\documentclass[reprint, superscriptaddress, secnumarabic, amssymb, nobibnotes, aps, prl]{revtex4-1}

\usepackage{graphicx}
\usepackage{epstopdf}
\usepackage[T1]{fontenc}
\usepackage{amsbsy}
\usepackage{gensymb}
\usepackage[T1]{fontenc}
\usepackage{amsmath}
\usepackage{amssymb}
\usepackage{bbm}
\usepackage{braket}
\usepackage{xcolor}
\allowdisplaybreaks
\usepackage{graphicx}
\usepackage[colorlinks=true]{hyperref}  
\hypersetup{
    bookmarks=true,         
    unicode=false,          
    pdftoolbar=true,        
    pdfmenubar=true,        
    pdffitwindow=false,     
    pdfstartview={FitH},    
    pdftitle={Superconducting properties of Re$_{2}$X (X = Hf, Zr)},    
    pdfauthor={Manasi Mandal},      
    pdfsubject={},   
    pdfcreator={},   
    pdfproducer={}, 
    pdfkeywords={} {} {}, 
    pdfnewwindow=true,      
    colorlinks=true,       
    linkcolor=blue, 
    citecolor=blue,        
    filecolor=magenta,      
    urlcolor=blue           
} 

\usepackage[normalem]{ulem}

\newcommand{\equref}[1]{Eq.~(\ref{#1})}

\newcommand{\figref}[1]{Fig.~\ref{#1}}

\newcommand{\tableref}[1]{Table~\ref{#1}}

\renewcommand{\approx}{\simeq}

\begin{document}

\title{\textrm{Study of the superconducting ground state of topological superconducting candidates  Ti$_{3}$X (X = Ir, Sb)}}
\author{Manasi~Mandal}
\affiliation{Department of Physics, Indian Institute of Science Education and Research Bhopal, Bhopal, 462066, India}
\author{Sajilesh~K.~P.}
\affiliation{Department of Physics, Indian Institute of Science Education and Research Bhopal, Bhopal, 462066, India}
\author{Rajeswari~Roy~Chowdhury}
\affiliation{Department of Physics, Indian Institute of Science Education and Research Bhopal, Bhopal, 462066, India}
\author{D.~Singh}
\affiliation{ISIS Facility, STFC Rutherford Appleton Laboratory, Harwell Science and Innovation Campus, Oxfordshire, OX11 0QX, UK}
\author{P.~K.~Biswas}
\affiliation{ISIS Facility, STFC Rutherford Appleton Laboratory, Harwell Science and Innovation Campus, Oxfordshire, OX11 0QX, UK}
\author{A.~D.~Hillier}
\affiliation{ISIS Facility, STFC Rutherford Appleton Laboratory, Harwell Science and Innovation Campus, Oxfordshire, OX11 0QX, UK}
\author{R.~P.~Singh}
\email[]{rpsingh@iiserb.ac.in}
\affiliation{Department of Physics, Indian Institute of Science Education and Research Bhopal, Bhopal, 462066, India}

\date{\today}

\begin{abstract}

The topologically non-trivial band structure of A15 compounds has drawn attention owing to the possible realization of topological superconductivity. Here, we report a microscopic investigation of the superconducting ground state in A15 compound Ti$_{3}$X (X = Ir, Sb) by muon spectroscopy measurements. Zero field muon measurements have shown that time-reversal symmetry is preserved in these materials. Furthermore, specific heat and a transverse field muon spectroscopy measurement rule out any possibility to have a nodal or anisotropic superconducting gap, revealing a conventional s-wave nature in the superconducting ground state. This work classifies A15 compound Ti$_{3}$X (X = Ir, Sb) as a time-reversal preserved topological superconductor.

\end{abstract}
\maketitle

\section{INTRODUCTION}
Spin-orbit coupling induced topological insulators (TI) are exceptional due to their unique property and incredible fundamental and applications-based intrigue. \cite{TI1}. Moreover, topological superconductors (TSC) featuring fully gapped bulk and gapless surface states have added to this search of new materials \cite{TI2,TS1}. Theoretical studies have proposed that topological superconductivity can be induced on a topological insulator surface in proximity to superconductors where the Dirac cone type surface states are forced to favor a $p_{x} + ip_{y}$ pairing \cite{TS2, TS3}. Several attempts have been made to realize superconductivity in TIs either by heterostructure fabrication or carrier doped topological insulators like Cu$_{x}$Bi$_{2}$Se$_{3}$ where features suggesting the existence of Majorana excitations can be observed \cite{biSe}. However, the possible complexity at the interfaces means extending the idea to search for conventional s-wave superconductors with topological surface states.
\\

Recent studies suggest that bulk superconductors can also possess a topologically nontrivial band structure \cite{TSC}. One of the main features behind these systems nontrivial behavior is the presence of strong spin-orbit coupling (SOC) that is strong enough to invert the band structure and leads to complete spin-momentum locking in the surface state \cite{TI2}. Among these, A15 superconductors, a well-known family of metal-based compounds, are promising candidates to realize topological superconductivity. Reports suggest, the standard crystal structure symmetry of A15 compounds, along with SOC, creates gapped crossing near the Fermi level. The unnoticed topological surface states near the Fermi surface of A15 compounds Ta$ _{3} $Sb, Ta$ _{3} $Sn, and Ta$ _{3} $Pb was unveiled in the recent theoretical calculation, indicating the potential candidacy to host topological superconductivity \cite{A151, A152}. The topological bulk band structures of these compounds can be characterized by nontrivial Z$_{2}$ invariants, with topological surface bands appear near the Fermi surface as dictated by the bulk boundary correspondence. Ta$_{3}$Sb houses an eight-fold degenerate Dirac point close to the Fermi level at symmetry point \cite{A151}. The first principle report on nonmagnetic A15 superconductors with high SOC has shown giant intrinsic spin Hall conductivities due to its many gapped Dirac crossings resulting in giant spin Berry curvature and correspondingly enormous intrinsic spin Hall effect \cite{A152}. Interestingly, the giant spin Hall conductivity in these materials has shown potential application in fault-tolerant topological quantum computation. 
Superconductivity in topological materials reported hosting various unconventional properties. For example, Cu$ _{x}$Bi$ _{2}$Se$ _{3}$ shows odd pairing \cite{CuBiSe}, rotational symmetry breaking, time-reversal symmetry breaking and superconducting topological surface state was observed in Sr$ _{x}$Bi$ _{2}$Se$ _{3}$ \cite{SrBiSe, SrBiSe2}, and PbTaSe$ _{2}$ \cite{PdTaSe}, respectively. Due to the limited number of topological superconducting materials, the pairing mechanism is not clearly understood. In this paper, we report detailed $ \mu $SR investigation of the superconducting ground state of topological superconducting materials Ti$ _{3} $Ir and Ti$ _{3} $Sb coupled with bulk measurements magnetization, heat capacity, and resistivity.\\

Zero field muon spin rotation and relaxation measurement ($ \mu $SR) is an excellent means to study unconventional superconductivity. Spin-relaxation experiments in zero field (ZF) can detect spontaneous magnetization that can be associated with spin-triplet superconductivity \cite{SrRuo1,SrRuo2,UPt1,UPt2,UPt3}. Also, $ \mu $SR can measure the field distribution across the sample and can give magnetic penetration depth to a high degree of accuracy. Temperature dependence of the magnetic penetration depth can unambiguously determine multiband superconductivity, line or point nodes, as well as anisotropy in the order parameter.

\section{EXPERIMENTAL AND CALCULATION DETAILS}
The compounds investigated were prepared from stoichiometric mixtures of Ti (99.99 $ \% $) powder and Ir (99.97 $\% $) or Sb (99.9 $\% $) pieces in a nominal ratio of 3:1 by standard arc melting method with negligible mass loss. The samples were flipped and remelted several times to improve the chemical homogeneity. Later, the compounds were annealed at 900\degree C for one week. The phase purity and crystal structure of the prepared samples were examined by PANalytical X$^{,}$pert Pro diffractometer equipped with Cu $K_{\alpha}$ radiation ($\lambda$ = 1.54056 $\text{\AA}$). Room temperature Rietveld refinement was performed using Full Prof Suite Software. We measured DC susceptibility by Superconducting Quantum Interference Device (SQUID MPMS, Quantum Design) under different applied fields to study the superconducting properties. Specific heat measurements were performed by the two tau time-relaxation method in zero field in the Physical Property Measurement System (PPMS, Quantum Design, Inc.). We carried out the $\mu$SR measurements on both the samples at the MuSR instrument of the ISIS neutron and muon source in STFC Rutherford Appleton Laboratory, United Kingdom. Zero-ﬁeld muon spin relaxation (ZF-$\mu$SR) was performed in the temperature range 0.3 - 9 K, and the transverse-ﬁeld muon spin rotation (TF-$\mu$SR) measurements were performed well above the lower critical field H$_{C1}$(0). 

\section{RESULTS AND DICUSSION}

\subsection{Sample characterization}
\begin{figure}
\includegraphics[width=1.0\columnwidth]{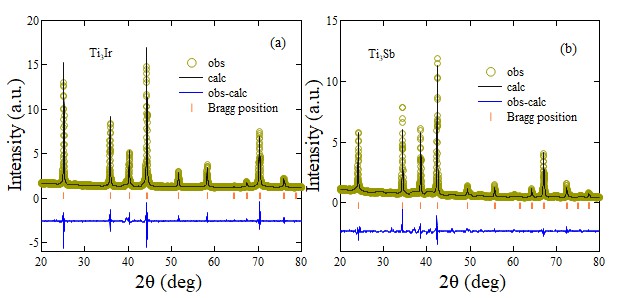}
\caption {\label{Fig1.jpg:XRD}Powder XRD pattern for (a) Ti$_{3}$Ir and (b) Ti$_{3}$Sb. The greenish circles and the solid black line correspond to the experimental data and calculated pattern, respectively. The red vertical bars give the Bragg positions, whereas the blue line at the bottom of the plot displays the difference between the experimental and calculated patterns.}
\end{figure}

Room temperature powder X-ray diffraction (XRD) analysis revealed that the samples are indexed well by cubic Cr$_{3}$Si-type structure with the space group Pm-3n (no. 223). The Rietveld refined powder XRD patterns of Ti$_{3}$X (X = Ir, Sb) are shown in (\figref{Fig1.jpg:XRD}). We have summarized the lattice parameters in the \tableref{tbl:lattice parameters}, which is consistent with the previous report \cite{TI, TSb1}.

\begin{table}[htbp!]
\caption{Refined lattice parameters of Ti$_{3}$X (X = Ir, Sb)}
\label{tbl:lattice parameters}
\setlength{\tabcolsep}{7pt}
\begin{center}
\begin{tabular}[b]{l c c c }\hline 
 Parameters & unit & Ti$_{3}$Ir & Ti$_{3}$Sb\\
\hline\hline
a &$\text{\AA}$ & 5.008(2) & 5.218(1)  \\
V$_{cell}$& $\text{\AA}^{3}$&  125.89 & 142.10\\
\hline
\end{tabular}
\par\medskip\footnotesize
\end{center}

\begin{center}
\begin{tabular}[b]{l c c c c}\hline 
Atom & Wyckoff position & x & y & z\\
\hline
\\[0.1ex]                               
Ir/Sb & 2a & 0.0 & 0.0 & 0.0 \\
Ti & 6c & 0.25 & 0.0 & 0.50\\
\\[0.1ex]
\hline
\end{tabular}
\par\medskip\footnotesize
\end{center}
\end{table} 

\subsection{Normal and Superconducting State properties}

\begin{figure}[htbp!]
\includegraphics[width=1.0\columnwidth]{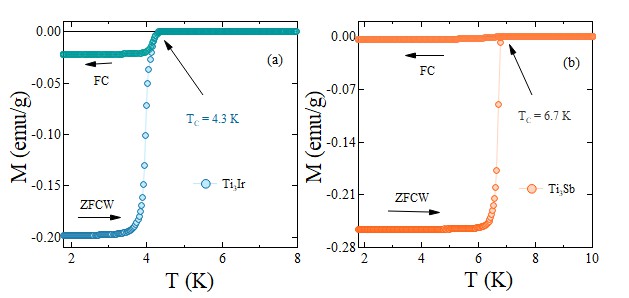}
\caption {\label{Fig2.jpg:ZFC} Temperature dependence of the magnetic susceptibility at 10 mT applied field exhibits the superconducting transition temperature.}
\end{figure}

We performed magnetization measurements in both zero-field cooled (ZFCW) and field cooled (FC) orders with a low applied field of 10 mT.  The ZFC and FC curves indicate the type-II nature of Ti$_{3}$X (X = Ir, Sb)  with superconducting transition temperature, T$_{C}$ = 4.3 $\pm$ 0.1 K and 6.7 $\pm$ 0.1 K, respectively, as shown in (\figref{Fig2.jpg:ZFC}). This result is compatible with the earlier report \cite{TSb2}.

\begin{figure}[htbp!]
\includegraphics[width=1.0\columnwidth]{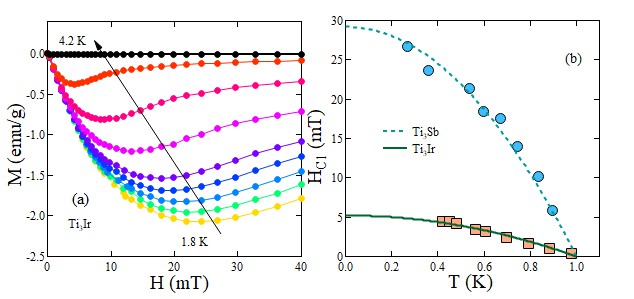}
\caption {\label{Fig3.jpg:HC1} (a) Variation of Magnetic moments with external field at different temperature. (c) Temperature dependence of the lower critical field H$_{C1}$ was fitted using Ginzburg-Landau relation.}
\end{figure}

We have also measured the magnetization with the variation of the magnetic field, M(H) at different temperatures, as shown in (\figref{Fig3.jpg:HC1})(a). The lower critical field H$_{C1}$(0) was calculated from the low field magnetization curves M(H) by fitting the data in accordance with Ginzburg-Landau (GL) Eq.\eqref{eqn3:HC1}:

\begin{equation}
H_{C1}(T)=H_{C1}(0)\left(1-\left(\frac{T}{T_{C}}\right)^{2}\right)
\label{eqn3:HC1}
\end{equation}

\begin{figure}
\includegraphics[width=1.0\columnwidth]{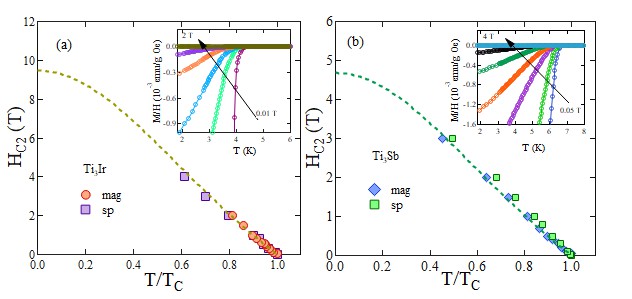}
\caption {\label{Fig4.jpg:HC2} Magnetization measurements for (a) Ti$_{3}$Ir and (b) Ti$_{3}$Sb were done at different fields.  H$_{C2}$(0) was calculated via GL- formula. }
\end{figure}

The calculated values of H$_{C1}$(0) are 5.2 $\pm$ 0.1 mT and 29.2 $\pm$ 0.4 mT for Ti$_{3}$X (X = Ir, Sb respectively)(b). To calculate the second-order transition field H$_{C2}$(0), magnetization as well as specific heat measurements were done in different applied magnetic fields up to 4.0 T, as shown in \figref{Fig4.jpg:HC2}. We have estimated the value of upper critical field H$_{C2}$(0) by using GL formula 
\begin{equation}
H_{C2}(T) = H_{C2}(0)\frac{(1-(\frac{T}{T_{C}})^{2})}{(1+(\frac{T}{T_{C}})^{2})}
\end{equation}
  
The calculated values of H$_{C2}$(0) are 9.6 $\pm$ 0.1 T for Ti$_{3}$Ir and 4.7 $\pm$ 0.2 T for Ti$_{3}$Sb. The Pauli limiting field within the BCS theory is expressed as H$_{C2}^{p}$(0) = 1.83 T$_{C}$, which gives H$_{C2}^{p}$(0) = 7.9 $\pm$ 0.2 and 12.3 $\pm$ 0.3 T for Ti$_{3}$X (X = Ir, Sb respectively). It is  very interesting that, H$_{C2}$(0) has exceeded the Pauli limit in case of Ti$_{3}$Ir, indicating a possible unconventional pairing.

We have calculated superconducting characteristics parameters, GL coherence length $\xi_{GL}$(0) and penetration depth $\lambda_{GL}$(0) by using the relations \equref{eqn6:coherence}, and \equref{eqn8:lamda}.
\begin{equation}
H_{C2}(0) = \frac{\Phi_{0}}{2\pi\xi_{GL}^{2}}
\label{eqn6:coherence}
\end{equation}
\begin{equation}
H_{C1}(0) = \frac{\Phi_{0}}{4\pi\lambda_{GL}^2(0)}\left( ln \frac{\lambda_{GL}(0)}{\xi_{GL}(0)}+0.497\right)  
\label{eqn8:lamda}
\end{equation}
where $\Phi_{0}$ (= 2.07 $\times$10$^{-15}$ T m$^{2}$) is the magnetic flux quantum \cite{tin}. The estimated values of $\xi_{GL}$(0) and $\lambda_{GL}$(0) were 58.6 $\pm$ 0.5 $\text{\AA}$ and 3851 $\pm$ 8 $\text{\AA}$ for Ti$_{3}$Ir and 83.9 $\pm$ 0.3 $\text{\AA}$ and 1361 $\pm$ 3 $\text{\AA}$ for Ti$_{3}$Sb. The value of GL parameter defined as $k_{GL}$ = $\frac{\lambda_{GL}(0)}{\xi_{GL}(0)}$ = 65.7 $\pm$ 0.1 and 16.2 $\pm$ 0.2 for Ti$_{3}$X (X =Ir, Sb) indicate the compounds as strong type-II superconductor. Thermodynamic critical field H$_{C}$ is estimated around 0.11  $\pm$ 0.02 T and 0.22  $\pm$ 0.03 T for Ti$_{3}$X (X = Ir, Sb) from the relation $H_{C1}(0)H_{C2}(0)$ = $H_{C}^2lnk_{GL}$.

\begin{figure}[htbp!]
\includegraphics[width=1.0\columnwidth]{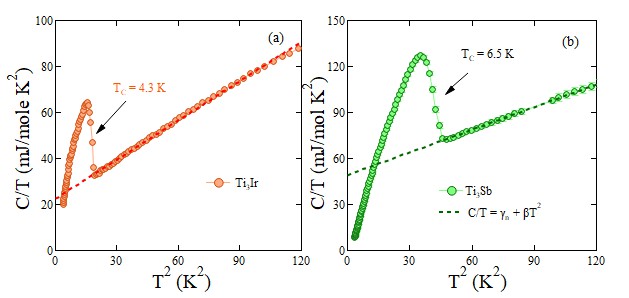}
\caption {\label{Fig5.jpg:SP} The variation of C/T with T$^{2}$ in zero fields where the pronounced jump indicate the superconducting transition. The dotted line is the fit in the normal state.}
\end{figure}

The samples were further characterized by heat capacity measurements at zero fields. A pronounced jump in heat capacity data indicates the superconducting transition, as observed in \figref{Fig5.jpg:SP}. To analyze the phonon properties and electronic density of states of the samples, we fitted the normal-state specific heat data by using the relation $\frac{C}{T}$ = $\gamma_{n}+\beta T^{2}$ where $\gamma_{n}$ is Sommerfeld coefficient and $\beta$ is Debye constant. The extrapolation of normal state behavior to the T $\to$ 0 limits allows the determination of normal state coefficients as $\gamma_{n}$ = 22.5 $\pm$ 0.1 mJ/mol K and $\beta$= 0.57 $\pm$ 0.02 mJ/mol K$
^{2}$ for Ti$_{3}$Ir and $\gamma_{n}$ = 49.2 $\pm$ 0.3 mJ/mol K and $\beta$= 0.49 $\pm$ 0.03 mJ/mol K$
^{2}$ for Ti$_{3}$Sb.

\begin{figure}[htbp!]
\includegraphics[width=1.0\columnwidth]{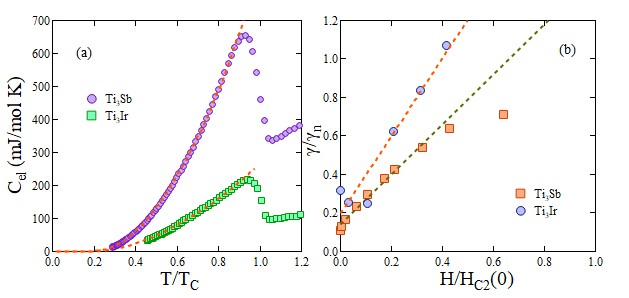}
\caption {\label{Fig6.jpg} (a) The specific-heat data in superconducting region fitted for single-gap s-wave model. (b) $\gamma$ and $\textit{H}$ were plotted versus each other after normalizing by $\gamma_{n}$ and H$_{C2}$(0) and fitted with straight line.}
\end{figure}

According to the Debye model Debye temperature $\theta_{D}$ is correlated to $\beta$ by the relation $\theta_{D}$ = $\left(\frac{12\pi^{4}RN}{5\beta}\right)^{\frac{1}{3}}$ where N (= 4) is the number of atoms per formula unit, R is the molar gas constant (= 8.314 J mol$^{-1}$ K$^{-1}$). The evaluated $\theta_{D}$ value was 239 $\pm$ 1 K and 25 $\pm$ 1 K for Ti$_{3}$X (X = Ir, Sb). We have used the value of $\gamma_{n}$ to extract the density of states $D_{C}(E_{F})$ at the Fermi level from the relation  $\gamma_{n}$ = $\left(\frac{\pi^{2}k_{B}^{2}}{3}\right)D_{C}(E_{f})$, where k$_{B}$ $\approx$ 1.38 $\times$ 10$^{-23}$ J K$^{-1}$. The approximated values are 9.5 $\pm$ 0.2 and 20.9 $\pm$ 0.1 $\frac{states}{eV f.u}$ for Ti$_{3}$X (X = Ir, Sb respectively). McMillan equation,  $\lambda_{e-ph}$ = $\frac{1.04+\mu^{*}ln(\theta_{D}/1.45T_{C})}{(1-0.62\mu^{*})ln(\theta_{D}/1.45T_{C})-1.04 }$ was used to calculate the strength of the attractive interaction between electron and phonon ($\lambda_{e-ph}$) \cite{McMillan}. The calculated values of $\lambda_{e-ph}$ = 0.65 $\pm$ 0.02 for Ti$_{3}$Ir and 0.74 $\pm$ 0.01 for Ti$_{3}$Sb suggesting that Ti$_{3}$X is a moderately coupled superconductor similar to Re$_{6}$X (X = Zr, Hf) \cite{Zr, ReHf}.
To find the behavior of the superconducting ground state, electronic specific heat was extracted from total specific heat data by the relation $C_{el}$ = $C -\beta_{3}T^{3}$. The data was well fitted with the BCS theory (\figref{Fig6.jpg}) \cite{BCS}. This yielded the value of the superconducting energy gap as $\frac{\Delta(0)}{k_{B}T_{C}}$ = 1.78 $\pm$ 0.04 for Ti$_{3}$Ir and 1.90 $\pm$ 0.03 for Ti$_{3}$Sb which is close to the BCS predicted value.\\
Specific heat measurements was performed at different applied field to determine the field dependence of Sommerfeld coefficient, $\gamma$(H) using the relation C$_{el}$/T versus T data with the equation $\frac{C_{el}}{T}$ = $\gamma + \frac{A}{T}exp\left(\frac{-bT_{C}}{T}\right)$ \cite{sas}. Field dependence of Sommerfeld coefficient $\gamma$ is shown in \figref{Fig6.jpg}(b), where $\gamma$ and H are normalized by $\gamma_{n}$ and H$_{C2}$(0), respectively. Here we can observe $\gamma$ increases linearly with H in the lower field region as shown in the \figref{Fig6.jpg}(b) indicating fully gapped superconducting energy gap.

\begin{figure}[htbp!]
\includegraphics[width=1.0\columnwidth]{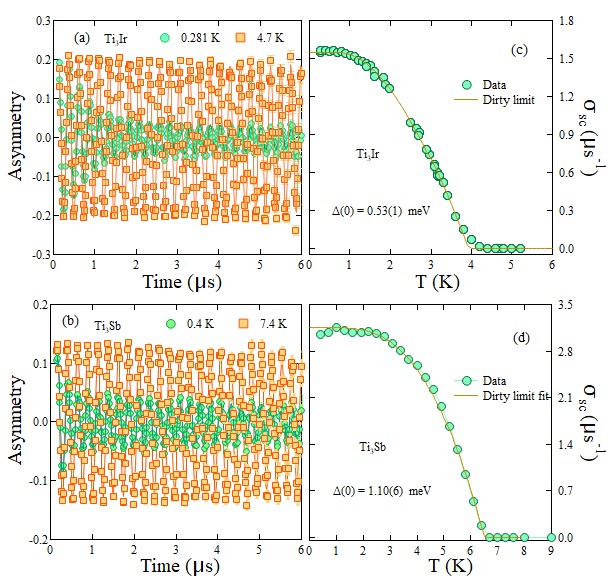}
\caption {\label{Fig7.jpg:TF} TF-$\mu$SR spectra above and below T$_{C}$ at 30 mT is shown in (a) for Ti$_{3}$Ir and in (b) for Ti$_{3}$Sb.  The spectra below T$_{C}$  decays faster for both samples indicating vortex formation. (c) and  (d) The variation of depolarisation across flux line lattice. The solid line is the fit to s-wave model in dirty limit. }
\end{figure}

We have further studied Ti$_{3}$X compound by muon spin rotation and relaxation measurements to explore the nature of the superconducting gap and symmetry of the order parameter. TF-$\mu$SR  measurements were done in a field-cooled mode (30 mT) down to 0.28 K for Ti$_{3}$Ir and 0.4 K for Ti$_{3}$Sb. TF-$\mu$SR precessional signals of both the samples at temperatures above and below T$_{C}$ is shown in \figref{Fig7.jpg:TF}. The normal state data shows a homogeneous field distribution throughout the time scale evident from the signal's non-decaying nature, whereas the superconducting state shows a decaying signal due to nonuniform field distribution in the flux line lattice (FLL) state. The time-domain signals were best fitted by the decaying Gaussian oscillatory function.
\begin{eqnarray}
G_{\mathrm{TF}}(t) &=&
A_{\mathrm{1}}\mathrm{exp}\left(\frac{-\sigma_{\mathrm{}}^{2}t^{2}}{2}\right)\mathrm{cos}(\gamma_{\mu}Bt+\phi)\nonumber\\&+&A_{bg}\mathrm{cos}(\gamma_{\mu}B_{bg}t+\phi)
\label{eqn1}
\end{eqnarray}
where $\phi$ is the initial phase of the muon spin polarization with respect to detector, $A_{\mathrm{1}}$ is the asymmetry, $B$ is mean-field of the Gaussian distribution, $\sigma_{\mathrm{}}$ is relaxation rate, and $\gamma_{\mu}$/2$\pi$ = 135.5 MHz/T is muon gyromagnetic ratio.  An additional oscillatory term consists of the asymmetry A$_{bg}$, and the field B$_{bg}$ were also added to compensate for the background contribution. 

$\sigma$ is the total depolarization arising due to the field variation across the flux line lattice $\sigma_{\mathrm{sc}}$, and the nuclear dipolar moments, $\sigma_{\mathrm{N}}$. We have extracted $\sigma_{\mathrm{sc}}$ by the relation $\sigma^{2}$ = $\sigma_{\mathrm{sc}}^{2}+\sigma_{\mathrm{N}}^{2}$. The variation of $\sigma_{\mathrm{sc}}$ with temperature was calculated by the second-moment method \figref{Fig7.jpg:TF} (c) and (d).  It is clearly seen that below $\approx$ $T_{c}$/3, the temperature dependence of $\sigma_{sc}$ is seen nearly constant for both the compound. This indicates the absence of nodes in the energy gap at the Fermi surface. We have fitted the data in the dirty limit within the local London approximation for a BCS superconductor. The corresponding expression is given by the equation: 
\begin{equation}
\frac{\sigma_{sc}^{-2}(T)}{\sigma_{sc}^{-2}(0)} = \frac{\Delta(T)}{\Delta(0)}\mathrm{tanh}\left[\frac{\Delta(T)}{2k_{B}T}\right] ,
\label{eqn5}
\end{equation}
where  $\Delta(0)$ is the energy gap magnitude at zero temperature and $\Delta(T)/\Delta(0) = \tanh\{1.82(1.018({T_{c}/T}-1))^{0.51}\}$ is the BCS approximation for the temperature variation of the energy gap. 
The fit yields $ \Delta_{0} $ = 0.53 $ \pm $ 0.01 meV and 1.10 $\pm$ 0.06 meV for Ti$_{3}$X (X = Ir, Sb respectively). The calculated value of normalised superconducting energy gap  $ \frac{\Delta(0)}{k_{B} T_{c}} $ is 1.54 $\pm$ 0.02 for Ti$_{3}$Ir and 1.90  $\pm$ 0.02 for Ti$_{3}$Sb, whereas the BCS predicted value is 1.76. 

We have calculated the London penetration depth $\lambda (0)$ = 2630  $\pm$ 20 \text{\AA} and 1845  $\pm$ 15 \text{\AA} for Ti$_{3}$X (X = Ir, Sb respectively) by the relation
\begin{equation}
\frac{\sigma_{sc}^{-2}(0)}{\gamma_{\mu}^{2}} = 0.00371* \frac{\phi_{0}^{2}}{{\lambda}^{4}(0)} 
\label{eqn6}
\end{equation}
These values are consistent with magnetization measurements. The ratio $\frac{T_{C}}{\lambda^{-2}(0)}$ is estimated as 0.30  $\pm$ 0.02 for Ti$_{3}$Ir and 0.22  $\pm$ 0.02 for Ti$_{3}$Sb which is ten times greater than the BCS limit 0.00025 $\sim$
0.015 \cite{uemura2, uemura3}.

\begin{figure}[htbp!]
\includegraphics[width=1.0\columnwidth]{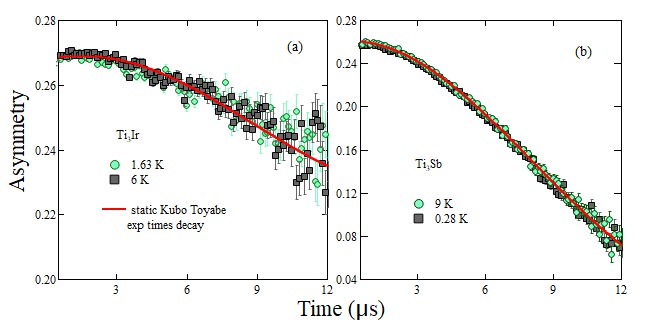}
\caption {\label{Fig8.jpg} ZF-$\mu$SR spectra above and below T$_{C}$  (a) Ti$_{3}$Ir and (b) Ti$_{3}$Sb.  The solid line is the fit to static Kubo Toyabe exponential times decay function. }
\end{figure}

\begin{figure}[h!]
\includegraphics[width=1.0\columnwidth]{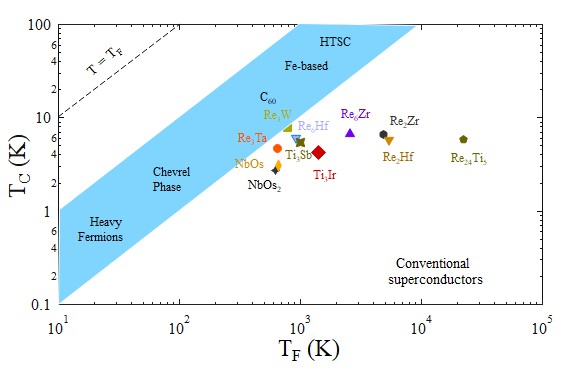}
\caption {\label{Fig9.jpg:Uemura} Uemura plot placed Ti$_{3}$X (X = Ir, Sb) along with some other unconventional materials.}
\end{figure}

The measured ZF-$\mu$SR spectra below and above spectra are shown in \figref{Fig8.jpg}.  The asymmetry spectra were best fitted by static Kubo Toyabe function multiplied by an exponential decay \cite{kubo}. There is no change in asymmetry spectra for the samples at temperatures above and below the superconducting transition. This suggests the absence of a spontaneous magnetic field below the superconducting transition temperature. This clearly indicates that the time-reversal symmetry is preserved in both the samples within the detection limit of $\mu$SR. Hence this work upholds the A15 compounds into time reversal symmetry preserved topological superconductors similar to rocksalt carbides NbC and TaC \cite{NbC}. 
We have calculated the Fermi temperature (T$_{F}$) for the samples in the dirty limit by simultaneously solving a set of five equations. A summary of the estimated parameters of both compounds is given in \tableref{tbl:parameters}. According to Uemura Classification scheme \cite{uemura1, uemura2, uemura3, uemura4}, the ratio $\frac{T_{C}}{T_{F}}$ for unconventional superconductors falls in the range 0.01$ \leq $ $ \frac{T_{c}}{T_{F}} $ $ \leq $0.1, while for conventional superconductors, $ \frac{T_{c}}{T_{F}} \leq$ 0.001. The estimated values of T$_{F}$ are 1399  $\pm$ 33 K and 994  $\pm$ 17 K, which gives the ratio $\frac{T_{C}}{T_{F}}$ as 0.003 for Ti$_{3}$Ir and 0.0055 for Ti$_{3}$Sb places the compounds out of the band of an unconventional family but close to many other exotic superconductors \cite{Zr, ReTi, ReTi2, NbOs, NbOs2, uemura5, uemura6, uemura7, uemura8} (\figref{Fig9.jpg:Uemura}).

\begin{table}
\caption{Normal and superconducting parameters of Ti$_{3}$X (X = Ir, Sb)}
\label{tbl:parameters}
\setlength{\tabcolsep}{10pt}
\begin{center}
\begin{tabular}[b]{l c c c}\hline 
Parameters& unit& Ti$_{3}$Ir & Ti$_{3}$Sb \\
\hline
\\[0.2ex]                             
T$_{C}$ & K & 4.3 & 6.7  \\
H$_{C1}(0)$ & mT & 5.2 &   29.2 \\ 
H$_{C2}$(0) & T & 9.6 &   4.7 \\
H$_{C2}^{P}$(0) & T & 7.9&   12.3  \\
$\xi_{GL}$& \text{\AA}& 58.6 &   83.9 \\
$\lambda_{GL}$& \text{\AA}& 3851 &   1361 \\
$k_{GL}$& & 65.7 &   16.2\\
$\gamma_{n}$& mJ/mol K$^{2}$& 22.5 &  49.2 \\
$\theta_{D}$& K& 239&  251\\
$\frac{\Delta(0)}{k_{B}T_{C}}$ &   & 1.78 &  1.90  \\
$\lambda_{e-ph}$ &  & 0.65 &   0.74 \\
D$_{C}$(E$_{f}$) & states/eV f.u. & 9.5 & 20.9 \\
n & 10$^{27}$m$^{-3}$ & 15.0 & 16.5 \\
$\frac{m*}{m_{e}}$ &  & 18 & 27 \\
v$_{f}$  & 10$^{4}$ ms$^{-1}$ & 4.9 & 3.4\\
$\lambda_{e}$  &\text{\AA} & 3174 & 3707\\
$\frac{\xi_{0}}{\lambda_{e}}$ & & 2 & 2\\
T$_{F}$ & K & 1399 &994 \\
$\frac{T_{C}}{T_{F}}$  & & 0.003&  0.0055\\
\\[0.2ex]
\hline
\end{tabular}
\par\medskip\footnotesize
\end{center}
\end{table} 

\vspace*{-0.4cm}

\section{Conclusion}

We have studied topological candidate A15 compounds Ti$_{3}$Ir and Ti$_{3}$Sb elaborately by XRD, magnetization, specific heat, and muon spectroscopy measurements. Samples were in pure phase with no impurity peak. Magnetization measurements suggest both the compounds are type-II superconductors with transition temperatures 4.3 K and 6.7 K, respectively. The upper critical field for Ti$_{3}$Ir is higher than the Pauli limit indicating a possible unconventional pairing mechanism in the superconducting ground state. However, TF-$\mu$SR confirms the fully gaped superconducting order parameter in both the compounds. ZF-$\mu$SR measurements reveal that the time-reversal symmetry is preserved within the detection limit of $\mu$SR, categorizing these compounds into time reversal symmetry preserved topological superconducting candidates. To understand the role of the  nontrivial band topology/topological surface states on the superconducting ground state, it is vital to study more A15 superconductors.

\vspace*{-0.9cm}

\section{Acknowledgments} R.~P.~S.\ acknowledges the Science and Engineering Research Board, Government of India for the Core Research Grant CRG/2019/001028. Financial support from DST-FIST Project No. SR/FST/PSI-195/2014(C) is also thankfully acknowledged.

\end{document}